\documentclass[9pt,twocolumn,twoside]{pnas-new}

\templatetype{pnasresearcharticle} 
\setboolean{displaywatermark}{false}
\title{New loop expansion for the Random Magnetic Field Ising Ferromagnets at zero temperature}

\author[a]{Maria Chiara Angelini}
\author[b]{Carlo Lucibello} 
\author[a,c]{Giorgio Parisi}
\author[a,c]{Federico Ricci-Tersenghi}
\author[d]{Tommaso Rizzo}

\affil[a]{Dipartimento di Fisica, Sapienza University of Rome, P.~le A.~Moro 5, Rome 00185 Italy}
\affil[b]{Bocconi University, Milan Italy}
\affil[c]{INFN, Sezione di Roma1 \& CNR-Nanotec, Rome unit, Rome 00185 Italy}
\affil[d]{CNR, Istituto dei Sistemi Complessi (ISC), Rome 00185 Italy}

\significancestatement{The $\epsilon$-expansion around the upper critical dimension is a standard tool for studying critical phenomena of models defined on finite-dimensional lattices. However, it faces problems in describing strongly disordered models. Here we use a new loop expansion around the Bethe solution, an advanced mean-field theory, since it provides a complete description of the fluctuations that play an important role at low temperatures, especially at finite connectivity.\\
We study the random field Ising model, a prototypical strongly disordered model, via this new loop expansion. We find indeed new correcting terms in the correlation functions at the one-loop order, but these are subdominant with respect to those coming from the standard $\epsilon$-expansion that is then correct at this order.}

\correspondingauthor{\textsuperscript{2}To whom correspondence should be addressed. E-mail: giorgio.parisi@roma1.infn.it}

\begin{abstract} 
We apply to the Random Field Ising Model at zero temperature ($T\!=\!0$) the perturbative loop expansion around the Bethe solution.
A comparison with the standard $\epsilon$-expansion is made, highlighting the key differences that make the new expansion much more appropriate to correctly describe strongly disordered systems, 
especially those controlled by a $T\!=\!0$ RG fixed point. This new loop expansion produces an effective theory with cubic vertices. 
We compute the one-loop corrections due to cubic vertices, finding new terms that are absent in the $\epsilon$-expansion. 
However, these new terms are subdominant with respect to the standard, supersymmetric ones, therefore dimensional reduction is still valid at this order of the loop expansion.
\end{abstract}

\dates{This manuscript was compiled on \today}
\doi{\url{www.pnas.org/cgi/doi/10.1073/pnas.XXXXXXXXXX}}

\def\s{\sigma}
\def\f{\tau}
\def\<{\langle}
\def\>{\rangle}
\def\(({\left(}
\def\)){\right)}
\def\[[{\left[}
\def\]]{\right]}
\newcommand{\be}{\begin{equation}}
\newcommand{\ee}{\end{equation}}

\DeclareMathOperator*{\sign}{sign}

\begin{document}

\verticaladjustment{-2pt}

\maketitle
\thispagestyle{firststyle}
\ifthenelse{\boolean{shortarticle}}{\ifthenelse{\boolean{singlecolumn}}{\abscontentformatted}{\abscontent}}{}

\dropcap{T}he critical behavior of the Random magnetic Field Ising ferromagnetic Model (RFIM) has been the subject of intense scrutiny in the last forty years.  
It is one of the most studied problems in statistical mechanics; however, at present, there are some basic questions to which we are unable to answer. 
The physics of the problem is quite clear: {\sl naively}  one expects that in presence of a random magnetic field a ferromagnetic transition is still possible but its critical properties should be different from that of an ordinary ferromagnet \cite{BrayMoore}. 
These systems can be realized experimentally as diluted antiferromagnets in a field, assuming that these two classes of systems belong to the same universality class \cite{FishmanAntiferro, CardyAntiferro,BelangerReview}: 
a ferromagnetic transition is observed, however some of the measurements are quite difficult because of an exceptionally strong critical slowing down that forbids the system 
to thermalize well near and below the critical temperature. As far as the dynamical properties are concerned, the RFIM model shares many characteristics with other glassy systems \cite{Dynamics}. 

The first study of the RFIM criticality used the perturbative renormalization group based on the same diagrammatic expansion that has been crucial to computing the critical exponents of standard ferromagnets in dimensions $D=4-\epsilon$ \cite{perturbativeRFIM1, perturbativeRFIM2}.  
One way to treat the RFIM is to introduce an effective replicated $\phi_4$ model once the disorder has been integrated out. 
The diagrammatical rules are simple and we recall them in the Supporting Information (SI).
Surprisingly, considering as usual only the most divergent diagrams and under some assumptions (see below) \textit{Dimensional Reduction} (DR) holds, i.e.\ 
the critical properties (e.g. the critical exponents of this random system) are the same of a pure system in $D-2$ dimensions at each order of perturbation theory \cite{AharonyD-2,YoungDR,ParisiSourlas}.  
DR is an astonishing relation as far as it connects the properties of different systems defined in different dimensions. 
In certain cases, DR has been rigorously established \cite{BrydgesImbrie} \footnote{Dimensional reduction has been proved between lattice animals in $D$ dimensions and ferromagnets in constant imaginary magnetic field in $D-2$ dimensions and between the Langevin stochastic equations in $D+1$ dimensions (one time dimension and $D$ space dimensions) and Boltzmann statistical mechanics in $D$ dimensions.}. 

However, we know that DR cannot be always valid in the case of the RFIM. 
Indeed the lower critical dimension (i.e. the dimension where the ferromagnetic transition disappears) of the RFIM should be 2 plus the lower critical dimension of the standard Ising ferromagnetic model, which is 1.
Therefore, according to DR, there should be no transition in the RFIM in $D=3$: this is at variance with a simple argument \cite{ImryMa}, \cite{AizenmanWher}, showing that the lower critical dimension for the RFIM is $D=2$. 
The existence of a transition in $D=3$ clearly shows that DR fails for the RFIM at low dimensions.
The DR conjecture for the upper critical dimension instead (i.e. the dimension at and above which the critical exponents are those predicted by the mean field theory) is $D=6$ and is expected to be correct.

The origin of DR (and the possible cause of its failure) was identified when it was shown that the sum of the leading diagrams is related to the stationary points of the Landau-Ginzburg (LG) Hamiltonian
\be
\int d^D x\left(\frac12 (\partial \phi(x))^2+\frac12 \tau\phi(x)^2+\frac14 g \phi(x)^4 - h(x) \phi(x)\right),
\ee
where $h(x)$ is the random magnetic field. The stationary points of the LG effective Hamiltonian are the solutions of the equation
\be
-\Delta\phi(x)+\tau \phi(x)+g\phi(x)^3=h(x)
\ee
In the case where the LG effective Hamiltonian has only one stationary point (i.e.\ only one local minimum) 
the sum of the leading diagrams gives the correct results: this is true in the region $\tau\ge 0$ due to the convexity of the LG effective Hamiltonian.
When the free energy has multiple local minima, one should take the weighted sum of all the stationary points: in this case the results obtained just summing the leading diagrams are not correct. 
It can be shown that the LG effective Hamiltonian has many stationary points below the critical point ($\tau=0, g>0$) and the derivation of DR is not sound
\cite{Parisi1984,Lancaster1995}. 
However, these multiple solutions are not seen in the standard perturbative (in $g$) expansion, and one can formally derive the DR result in any dimension.

In which dimension does DR break down? One can conceive at least three scenarios.
\begin{enumerate}
\item Dimensional reduction gives a grossly wrong result also near $D=6$. In this scenario, the difference between the correct results and those coming from the $6-\epsilon$ expansion is already non-zero at order $\epsilon^k$ for some $k$. In other words, the results of the $\epsilon$ expansion are not correct even as a Taylor expansion in $\epsilon$: this would be not so surprising because some assumptions in the derivation are wrong.
\item  A less extreme suggestion is that this difference is exponentially small when $\epsilon \to 0$, e.g. it is of order $\exp (-A/\epsilon)$. 
This was argued in a non-fully conclusive way by Parisi, Dotsenko \cite{ParisiDotsenko}. 
The rationale for the suggestion is that since the difference is invisible in perturbation theory, 
it is reasonable to suppose that perturbation theory is still valid and that the difference is of order $\exp (-A/g_R)$, where $g_R$ is the renormalized dimensionless coupling constant. 
We know that at the critical point $g_R$ is proportional to $\epsilon$ and this leads to the aforementioned conclusion. In this scenario the perturbation theory is correct but there are non-perturbative exponentially small terms that have to be added to get the correct result. 
\item There is a critical dimension $D_{DR}$ such that for $D>D_{DR}$ dimensional reduction is correct and for $D<D_{DR}$ dimensional reduction fails. 
We can interpret this scenario by saying that there are two fixed points: the DR one and the one with broken DR. At $D=D_{DR}$ the DR fixed point becomes unstable and 
the stable one is the one with broken DR. This scenario has been strongly advocated by Tarjus and Tissier: 
using the  Non-Perturbative Functional Renormalization group, they found that DR is no more valid for $D<D_{DR}\simeq 5.1$ \cite{d5.1}
\end{enumerate}

Arguments in favor of each of three scenarios have been going on for a few decades. Some doubts on the correctness of LG based approaches come from the suggestion that the dominant fixed point of the RFIM is a $T=0$ one \cite{BrayMoore, FisherT0}.
This means that the RG flux starting at or below the critical temperature will evolve towards zero temperature. The analytic arguments are complemented by very large scale numerical studies confirming that the critical exponents at $T=0$ and at finite $T$ are the same \cite{Ogielski}. Nowadays most of the more accurate simulations are done at $T=0$ in the Ising case where there is only a quite minor effect of critical slowing down near the critical point \cite{simulazT0, d5Fytas2019}. In the LG approach, and at odds with the BL one, we cannot set $T=0$ from the onset but have to take the zero-temperature limit. 

In this paper, we argue that the first scenario is likely incorrect by showing that a different perturbative expansion leads to the same prediction as the $\epsilon$-expansion. As in functional renormalization group approaches, we don't treat perturbatively the non-free field term of the Hamiltonian. The start of our computation is not the LG Hamiltonian but the Hamiltonian of the the original RFIM model. We study the RFIM at $T=0$ using a new loop-expansion around the mean field Bethe solution, recently proposed in ref.~\cite{Mlayer_method}.
This expansion has some very interesting features that make it very promising:
\begin{itemize}
\item The mean field Bethe solution is obtained solving the model on a random regular graph, usually called Bethe lattice (BL), having the same coordination number of the physical finite dimensional lattice we are interested into. On the contrary, the standard LG effective Hamiltonian can be derived introducing an auxiliary lattice whose coordination number goes to infinity \cite{Mlayer_FC}. The finiteness of the coordination number introduces new features that produce crucial differences, as we will see later.
\item The RFIM on a BL can be studied in great detail and many of its properties can be analytically computed.
In standard mean field theory (that is valid on a fully connected lattice) the global magnetization is the only important variable and it satisfies a simple equation. On the contrary on the BL, the relevant quantity is the probability distribution of the local effective fields.
\item Computations can be done setting the temperature straight to 0. The leading order of the perturbation theory (i.e. the theory on the BL) at $T=0$  can be easily derived. Successive orders can be can be expressed as diagrams corresponding to physical generalized loops on the lattice and their contribution for large distances can be identified and computed. Their physical interpretation is clear.
\item Most importantly. we do not have to do assumptions on the uniqueness of the solution of some equations as it was implicitly done in the DR derivation.
\end{itemize}

In this paper, we shall see how this method is implemented. We stress that the case of the RFIM is just a particular application of the method that is much more general: a recent application shows that the mean-field hybrid transition of bootstrap percolation does not survive in finite dimension \cite{Rizzo} and it could also be used for spin glasses (with zero or non-zero magnetic field) and for the Anderson transition.

Let us consider for definiteness an explicit realization of the RFIM. The spins are located on the $D$-dimensional lattice $\mathbb{Z}^D$ with nearest neighbors interactions. The Hamiltonian is
\begin{equation}
H[\sigma,h]=-\frac12 \sum_{(i,j)} \sigma_i \sigma_j -\sum_i h_i^R \sigma_i,
\label{eq:H}
\end{equation}
where the sum is over nearest neighbors, $\sigma_i \in \{-1,+1\}$,  and the random magnetic fields $h_i^R$ are independent and identically distributed according to a Gaussian with  0 mean and variance $\sigma^2_h$.  At $T=0$, the only relevant configuration is the ground state of the system, that we call $\sigma^*$. 
In order to describe the critical behavior, we introduce the $T=0$ disconnected correlation $C_{ij}$ and response function $R_{ij}$:
\begin{equation}
C_{ij}  \equiv \overline{\s_i^*\s_j^*}\quad\qquad 
R_{ij}\equiv\frac{1}{2}\overline{1-\sigma^*_i\<\sigma_i\>_j}
\end{equation}
where the overline denotes the average over the random realizations of the fields $h^R$
and $\<\cdot\>_j$ denotes evaluation on the minimum energy configuration conditioned to having spin $j$ flipped with respect to the ground state, i.e. $\sigma_j=-\sigma^*_j$.
We note that $R_{ij}$ is the probability that site $i$ belongs to the lowest energy excitation of site $j$ (a droplet), therefore the sum over all sites $i$  yields the average droplet size, $\sum_i R_{ij}$.
We considered the above definition of $R_{ij}$ because it is convenient for numerical experiments, while alternative choices are possible and will be discussed below.
Both $C_{ij}$ and $R_{ij}$ are translational invariant, since they contain disorder averaged quantities.

\section*{Expanding around the Bethe solution}

The expansion proposed in ref.~\cite{Mlayer_method} is an expansion around the Bethe solution: by replicating the model $M$ times, and rewiring the $M$ copies,
one can show that the limit $M\to\infty$ gives the Bethe approximation, while the original model is recovered for $M=1$. A diagrammatic loop expansion can then be constructed expanding in powers of $1/M$, similarly to the standard perturbative expansion. We call this framework the $M$-layer or the BL approach.
At leading order, the correlation functions on the $M$-layer are strictly related to the ones found on a BL. The latter are easy to compute since in the thermodynamic limit the BL contains no loops of finite length.

Using the $M$-layer expansion, one finds that a generic correlation $G$ (either $C$ or $R$) 
between the lattice origin and a point $x\in Z^D$ on the lattice can be written, at the leading order, as
\begin{equation}
G(x)=\sum_{L=1,\infty}{\cal N}(x,L)\, G^{\text{\tiny{BL}}}(L)
\label{eq:G(x)}
\end{equation}
where ${\cal N}(x,L)$ is the number of non-backtracking paths going from the origin to $x$ of length $L$ on the original lattice and $G^{\text{\tiny{BL}}}(L)$ is the correlation function computed on the BL between two spins at distance $L$. The BL has the same connectivity $z=2D$ of the finite dimensional lattice and the same probability distribution for the random fields.

In the region of large $x$ and $L$ we find that in $D$ dimensions \cite{Rizzo19}
\begin{equation}
{\cal N}(x,L) \propto (2D-1)^L \exp\((-x^2/(4L)\)) L^{-D/2},
\end{equation}
where $x^2\equiv\lVert x\rVert^2$, and $\propto$ denote equality up to a constant. 
and we obtain for the Fourier transform of eq.~[\ref{eq:G(x)}] in the small momentum region 
\begin{equation}
\widetilde G(p)\propto\sum_{L=1,\infty}  (2D-1)^L  \exp(-L \, p^2) G^{\text{\tiny{BL}}}(L)\ .
\label{eq:G(p)}
\end{equation}
So we just need to compute how correlations behave on a BL at large distances.
As shown in the SI, at $T=0$ the crucial quantity that encodes all the information about two spins $\s_1$ and $\s_2$ at positions $x_1$ and $x_2$ is the dependence of the ground state energy on their values, after that we minimize over all the other spins.
For Ising spins the energy must be of the form
\begin{equation}
\mathcal H [\s_1,\s_2] = -h_1 \s_1  -h_2 \s_2 -J \s_1\s_2 + E.
\end{equation}
On the BL, we can derive a recursive equation for the joint probability distribution $P_L(h_1,h_2,J)$ for two spins at distance $L$, and obtain an explicit expression for large $L$  by imposing some consistency condition. The computation is  presented in the SI, and from here on we discuss the large $L$ behaviour, which is the relevant one at criticality. 
The result can be written in the form:
\begin{equation}
P_L(h_1,h_2,J)=Q_L(h_1,h_2)\delta(J)+a L\lambda^{L} \hat g(h_1) \hat g(h_2)F_L(J)
\label{eq:ansatz}
\end{equation} 
where $\lambda, \hat g, F_L$ and $Q_L$ depend implicitly on the external random field distribution and $F_L(J)=\rho L \exp (-\rho LJ)$ at the leading order in $L$. 
In spin-glass jargon, $\lambda$ is called the \emph{anomalous} eigenvalue and governs the decay of ferromagnetic susceptibilities along a chain in the BL \cite{Weigt96, Lucibello14}.   
The expression for the marginal probability $P_L(J)\equiv \int dh_1 dh_2\, P_L(h_1,h_2,J) $ is thus given by:
\begin{equation}
 P_L(J)=(1-aL\lambda^ L)\delta(J) +a\rho L^2\lambda^ L \exp (-\rho LJ)
 \label{Eq:PLJ}
\end{equation}
The coefficient $\rho$ can be computed exactly, as shown in the SI.
The result [\ref{Eq:PLJ}] is quite surprising: the quantity $J$ is either exactly 0 or is of order $1/L$ with a probability of order $L \lambda^ L$. Denoting with $\overline{\bullet}^L$ averages over $P_L$, we obtain
\begin{equation}
\overline{J}^L\propto \lambda^ L
\end{equation}
It can be shown that the response function $R_{12}$ receives contributions only from the event $J>|h_1|$, and can be easily computed. We obtain that the average response function on a BL behaves as the average effective coupling:
\begin{equation}
R^{\text{\tiny{BL}}}(L)\propto \lambda^ L
\label{eq:R^B}
\end{equation}
In a similar way, if we look to $P_L(h_1,h_2)=\int dJ P_L(h_1,h_2,J)$ we find that for large $L$
\begin{equation}
P_L(h_1,h_2)=P(h_1)P(h_2)+L \lambda^L f(h_1)f(h_2),
\end{equation}
where the function $f(h)$ is even and $P(h)$ is the Bethe distribution of cavity fields. One can see that the dominant contribution to the disconnected correlation on the BL at distance $L$
comes from the correlated fields $h_1$ and $h_2$. For this reason we find:
\begin{equation}
 C^{\text{\tiny{BL}}}(L)\propto L \lambda^ L
\label{eq:C^B}
\end{equation}
The behaviors for $R^{\text{\tiny{BL}}}$ and $C^{\text{\tiny{BL}}}$ correspond to the known ones on a line \cite{Flaviano,Lucibello14}.

We have found that for large $L$, $G^{\text{\tiny{BL}}}(L)\approx \mathcal{G}^{\text{\tiny{BL}}}(L) \lambda^{L}$, where $\cal{G}^{\text{\tiny{BL}}}$ is a polynomial in $L$.
On the BL, the correlation functions decrease exponentially and the critical point is located where their exponential 
decrease matches the exponential increase of the number of paths (resulting in a diverging susceptibility). In both the BL and in finite dimensional model at the zeroth-order of the $M$-layer construction, the critical point is located at
\begin{equation}
\lambda_c=\frac{1}{2D-1}\,.
\end{equation}
Near the critical point, starting from eq.~[\ref{eq:G(p)}] we can write
\begin{equation}
\widetilde G(p)\propto\int_0^\infty dL\ \exp\((-L (p^2 +\tau)\))  \mathcal{G}^{\text{\tiny{BL}}}(L)  \label{eq:propertime}
\end{equation}
where the sum over $L$ has been replaced by an integral and $\tau$ is the inverse of the correlation length $\xi$ and it is given by 
\begin{equation}
\tau\equiv -\log(\lambda(2D-1))
\end{equation}
As usual, $\tau=0$ at the critical point.
The representation in eq.~[\ref{eq:propertime}] is the equivalent of the proper time representation in a field theory context.

If we now put eqs.~[\ref{eq:R^B},\ref{eq:C^B}] in eq.~[\ref{eq:propertime}], we obtain the  leading order of the expansion of the correlation functions of a $D$-dimensional model  around the BL:
\begin{equation}
\label{BING}
\left.\begin{aligned}
 \widetilde{C}(p)\propto& \int_0^\infty dL\  \exp\((-L (p^2 +\tau)\)) L = \frac{1}{(p^2+\tau)^2}\\ 
 \widetilde{R}(p)\propto& \int_0^\infty dL\  \exp\((-L (p^2 +\tau)\)) = \frac{1}{p^2+\tau}
\end{aligned}\right.
\end{equation}

\subsection*{Comparison with the Dimensional Reduction}
\label{sec:comparison}

The formulae in eq.~[\ref{BING}] are the same of those coming from the LG effective Hamiltonian approach, with a few crucial differences though.
\begin{itemize}
\item In the LG approach, near the critical point and in the zero-loop approximation, the equations for the stationary points are linear: the unique solution is $\phi(x)\!=\!\int\!dy R(x-y) h(y)$, where $R(x)$ is the Fourier transform of $\widetilde{R}(p)$.
In the BL approach, near the critical point and in the zero-loop approximation, there is an infinite number of local minima (i.e.\ configurations whose energy does not decrease if a finite number of spins are flipped), but the only thermodynamically relevant configuration is the global minimum \cite{ManyStates}.
\item In the LG approach, the response function $R(x)$ does not depend on the field and it does not fluctuate: more precisely, all possible paths give the same contribution. 
In the case of the  BL approach, only an exponentially small number of paths gives a contribution to the response function: 
for large $L$ the probability of a given path to have a non zero $J$ and consequently to contribute to the response is exponentially small \cite{Flaviano}.
\end{itemize} 

The above differences become strikingly evident if we consider avalanches, a well-studied phenomenon in the RFIM.
We are interested in seeing the change in the magnetizations when we change by a finite amount the magnetic field in a given point. Denoting the original field at position $x$ by $h^*$, we define 
\begin{equation}
A(y,x;h)\equiv \langle\sigma(y)\rangle_{h^*+h} -\langle\sigma(y)\rangle_{h^*-h}
\end{equation}
where the label denotes the field at  $x$. The quantity $A(y,x;h)$ is 
 the variation of the magnetization at $y$ when we change the magnetic field at $x$ adding or subtracting a term $h$.
In the limit of small field avalanches are related to an alternative choice of the response function that we call $\hat{R}(y-x)$. It amounts to consider only (the density of) excitations with strictly zero energy cost. 
More precisely, we have:
\begin{equation}
\hat{R}(y-x)=\lim_{h\to 0}\overline{\frac{A(y,x;h)}{2 h}}.
\end{equation}
One can show that $\hat{R}(y-x)$ is proportional to the zero temperature limit of the two-point 
connected correlation function, with a factor proportional to the inverse temperature.

The avalanche size $S(x;h)$ is given by
\begin{equation}
S(x;h)=\frac{1}{2}\int d^Dy\, A(y,x;h)\,.
\end{equation}
For small $h$ we have 
\begin{equation}
\overline{S(x;h)}\approx  h\, \chi\equiv h \int dy\, \hat{R}(y-x)\,,
\end{equation}
with $\chi$ the susceptibility associated to the response function $\hat{R}(y-x)$.
The susceptibility  diverges as $1/\tau$ at the critical point both in the LG approach and on the Bethe lattice, but new features arise when we consider the probability distribution $P(S)$ of $S(x;h)$. Let us compare what happens in the two  approaches at the zeroth-order of the loop expansion.
\begin{itemize}
\item In the LG approach, since $\hat{R}$ is related to connected correlation functions we easily find that $S$ does not fluctuates: $P(S)=\delta\((S-\frac{c}{\tau}\))\,$
for some constant $c$. Therefore, we find:
$\overline{S}\propto \tau^{-1} ,\quad \overline{S^2}\propto \tau^{-2}$.
The median value of $S$ is divergent.
\item In the BL approach, following \cite{GarelMonthus}, one can argue that:
$P(S)\propto S^{-3/2} \exp(-S\tau^2 ).$
We thus find:
$\overline{S}\propto \tau^{-1}, \quad \overline{S^2}\propto \tau^{-3}$.
The median value of $S$ is finite and the divergence of $\overline{S}$ and $\overline{S^2}$  stems from rare events in the tail of the distribution.
\end{itemize}
The power law divergence of $\overline{S^2}$ is quite different in the two approaches: $\tau^{-2}$ in LG and $\tau^{-3}$ on the BL (more details in the SI).


\section*{1-st order in the Loop-expansion in the BL approach}

We have seen that at the zeroth-order the physical behavior is quite different in the BL and in the LG approaches, although the critical behavior of the correlation function is superficially similar: 
in the LG approach, anomalous large fluctuations do not exist, while on the BL everything is dominated by rare large fluctuations. 
The superficial similarity for the average two-point correlations disappears if we look to high-order correlation functions (responsible for avalanches).

At this point, it is not clear what happens when we consider the loop expansion in the BL approach. The natural question is whether this loop expansion produces the same results as in LG. We have two alternative scenarios:
\begin{itemize}
\item The difference in the high-order correlations that we have seen at the tree level (zeroth-order BL) contaminates the two-points correlations when the leading contributions 
coming from the loop are considered.
In this case, we would have additional terms at $T=0$ that are ignored in the LG approach. 
This would lead to the appearance of extra terms in the  $\epsilon$ expansion in $6-\epsilon$  and  DR should fail already in the $\epsilon$ expansion.
\item The difference in the high-order correlations do not produce leading discrepancies on the two-points correlation functions and the contribution of the loops is the same as in the LG approach. 
As a consequence, in an unexpected way, we would recover perturbative dimensional reduction in  $6-\epsilon$. 
\end{itemize}
One can present many hand-waving arguments in favor of the first or the second scenario. However, the proof of the pudding is in the eating. 
In the following we prove that at one loop the results of the BL and LG approaches are the same. 
This will be done presenting a computation ({\sl down to the metal}) of the one loop correction in the case of the BL. 

Roughly speaking the idea at the basis of the loop expansion around the BL is to start approximating, at least locally, the $D$-dimensional lattice with loopless (acyclical) graphs: these are Caley trees with self-consistent conditions at the boundary or Bethe lattices.
Of course, loops are present in the $D$-dimensional lattice and their effect is introduced perturbatively, by considering a sequence of BL with a finite number of loops. The expansion is similar to the virial expansion, where the complex interaction among infinitely many particles is decomposed in terms of simpler interactions between a finite number of particles.

\begin{figure}[t]
\begin{center}
\includegraphics[width=\columnwidth]{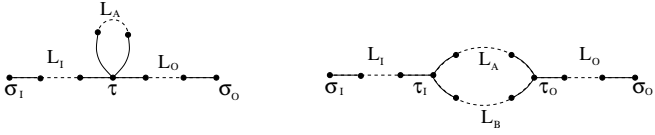}
\caption{One loop topological diagrams important for the first order expansion around the Bethe lattice: they can have vertices with four lines (left), or vertices with three lines (right).}
\label{Fig:1loopDiagram_phi4_phi3}
\end{center}
\end{figure}

The loop expansion around the BL is an expansion in topological diagrams.
The contribution of a given topological diagram can be written as the probability of finding such a topological diagram embedded in the $D$-dimensional lattice times 
the averaged value that the observable takes on that given structure when inserted in a loop-less and infinite  Bethe lattice. Only the \emph{topologically connected} part of the average of the observable has to be consider. As it is shown in ref.~\cite{Mlayer_method}, 
this \emph{connectivization} procedure practically corresponds to adding the value of the observable evaluated on each of the subgraphs that are obtained from the original structure 
by sequentially removing its lines times a factor $-1$ for each line removed.
The one loop contribution comes from the two diagrams shown in Fig.~\ref{Fig:1loopDiagram_phi4_phi3}.
They look similar to standard Feynman diagrams, however their physical interpretation is quite different. 
In standard Feynman diagrams the loops do not have a special meaning, here instead they have a geometrical meaning. Generalizing eq.~[\ref{eq:propertime}], the one loop contribution, in Fourier space and as a function of the incoming moment, can be written as
\begin{equation}
\widetilde{G}^{loop}(p)=\int_0^\infty d\vec L \ \tilde{\cal{N}}(p,{\vec L})\,\lambda^{\Sigma(\vec L)}\,{\cal G}^{\text{\tiny{BL}}}(\vec L)
\end{equation}
where $\vec L$ is a vector containing the lengths of each line in the  topological diagram,
the factor $\tilde{\cal{N}}(p,\vec L)$  accounts for the number of such topological diagrams, while  $\Sigma(\vec L)$ is the sum of all $L$'s, and $\lambda$ is the same eigenvalue on the BL as in the previous discussion.
The term ${\cal G}^{\text{\tiny{BL}}}(\vec L)$ is the generalization of $ {\cal G}^{\text{\tiny{BL}}}(L)$ at the zero-th order: it is the only term depending on the model and has to be carefully computed on the BL.
In the case of the two diagrams in Fig.~\ref{Fig:1loopDiagram_phi4_phi3}, we find:
\begin{itemize}
\item for the left diagram
\begin{align}
\tilde{\cal{N}}(p,\vec L)&\propto\frac{(2D-1)^{\Sigma(\vec L)}}{{\cal D}({\vec L})^{D/2}} \exp(-\left(L_I +L_O\right)p^2) \label{eq:Ntadpole}\\
  {\cal D}(\vec L)&=L_A, \qquad  \Sigma(\vec L)=L_I+L_O+L_A;\qquad\qquad\qquad \nonumber
\end{align}

\item for the right diagram
\begin{align}
\tilde{\cal{N}}(p,\vec L)&\propto\frac{(2D-1)^{\Sigma(\vec L)}}{{\cal D}({\vec L})^{D/2}}\exp\left(-\left(L_I +L_O+\frac{L_AL_B}{{\cal D}(\vec L)}\right)p^2\right), \label{eq:Nphi3}\\
{\cal D}(\vec L)&=L_A+L_B,\qquad \Sigma(\vec L)=L_I+L_O+L_A+L_B.\qquad \nonumber
\end{align}
\end{itemize}
Setting $\mathcal{G}(\vec L)=1$ we recover the conventional diagrams of the field theory approach in the cases of a $\phi^4$ interaction (left diagram) or $\phi^3$ interaction (right diagram) written in the Feynman proper time representation:
\begin{equation}
\left.\begin{aligned}
\widetilde{G}^{loop}_{\phi^4}(p)&=\frac{1}{(p^2+\tau)^2}\int d^Dq \:\frac{1}{q^2+\tau} \\
\widetilde{G}^{loop}_{\phi^3}(p)&=\frac{1}{(p^2+\tau)^2}\int d^Dq \:\frac{1}{q^2+\tau}\:\frac{1}{(p-q)^2+\tau}
\end{aligned}\right.
\end{equation}
where $\tau=-\log(\lambda(2D-1))$ as usual.
In fact, we can go backward from last expression containing integrals in momentum space to the previous one: for each line we have to use the representation
\begin{equation}
\frac{1}{p^2+\tau}=\int_0^{\infty} dL\  \exp\((-L(p^2+\tau)\))
\label{eq:singlePoleRepresentation}
\end{equation}
In this way, the integral over the loop  momentum $q$ becomes Gaussian: it can be readily done and we obtain the previous results, eqs. [\ref{eq:Ntadpole},\ref{eq:Nphi3}].

It is clear that the crucial point is the computation of the function ${\cal G}^{\text{\tiny{BL}}}(\vec L)$, since it contains all the information related to the theory we are considering. 
In the case of the standard LG approach, only the left diagram is present: a standard computation gives for the disconnected and the connected correlation functions
\begin{equation}
{\cal G}_C^{\text{\tiny{LG}}}(\vec L)=L_A (L_I+L_O) \qquad {\cal G}_R^{\text{\tiny{LG}}}(\vec L)=L_A
\end{equation}
and, using the representation 
\begin{equation}
\frac{1}{(p^2+\tau)^2}=\int_0^{\infty} dL \  L\,\exp\((-L(p^2+\tau)\))\,,
\label{eq:doublePoleRepresentation}
\end{equation}
we obtain the standard result where some lines have a single pole, $(p^2+\tau)^{-1}$, 
while others have a double pole $(p^2+\tau)^{-2}$. 
This representation can be derived also for higher orders of the perturbative expansion. 
The first perturbative proof \cite{YoungDR} of DR was based on the use of the identity for the diagrams:
${\cal G}_R^{\text{\tiny{LG}}}(\vec L)={\cal D}(\vec L).$ 
In this way, the denominator in $\tilde{\cal N}$ becomes ${\cal D}({\vec L})^{D/2-1}$ and the final expression for the diagrams is the same of a vanilla $\phi^4$ theory in dimensions $D-2$.

How to compute the factors ${\cal G}^{\text{\tiny{BL}}}(\vec L)$ in the BL approach?
We have to compute the average connected and disconnected correlations on a BL where we have the same local geometry ($z=2D$) plus a manually injected topological diagram. 
The final results can be summarized as follows 
\begin{itemize}
\item The left diagram gives the same type of contribution of the diagrams of LG reproducing DR.
\item In the region where either $L_A$ or $L_B$ is small the right diagram has a behavior quite similar to the left diagram. Nothing new comes from this diagram in this region.
\item The real interesting region for the right diagram is when all the $L$'s are large: 
this gives the relevant contribution at large distances (small momentum) discussed below.
\end{itemize}

\subsection*{Computation of the new diagram on the BL}

In order to compute factors ${\cal G}^{\text{\tiny{BL}}}(\vec L)$ on a BL, we have to go through a sequence of simple steps. Some of them are rather lengthy  yet straightforward. 

We consider a BL where we add a loop of the type of (Fig.~\ref{Fig:1loopDiagram_phi4_phi3}, right). Apart from the loop, the rest of the lattice is a standard BL with fixed connectivity $z=2D$: this means that variables $\s_I$ and $\s_O$ are the root of $z-1$ infinite tree-like branches, variables $\tau_I$ and $\tau_O$ are the root of $z-3$ tree-like branches, while the spins along the topological lines are the root of $z-2$ tree-like branches.

We are interested in computing the probability distribution of the random restricted Hamiltonian  $\mathcal H [\s_I,\s_O]$, i.e.\ the one we obtain after minimizing with respect to all the other variables.
This 2-spins Hamiltonian is obtained from the 4-spins Hamiltonian $\mathcal H [\s_I,\s_O,\tau_I,\tau_O]$, where the distances among are fixed to the values $L_I,L_O,L_A,L_B$ shown in Fig.~\ref{Fig:1loopDiagram_phi4_phi3}, by
\begin{equation}
\mathcal H [\s_I,\s_O]=\min_{\tau_I,\tau_O} \mathcal H [\s_I,\s_O,\tau_I,\tau_O]
\end{equation}
The 4-spins Hamiltonian can be computed by summing four statistically independent 2-spins Hamiltonian
and the cavity fields on $\s_I,\s_O,\tau_I,\tau_O$ coming from the infinite trees. A two-spin Hamiltonian for a line of length $L$ is described by two fields and a coupling, $(u_1, u_2, J)$, whose joint law for large $L$ can be written in the form
\begin{multline}
P_L(u_1,u_2,J)=P(u_1)P(u_2)\delta(J)+\\
+Q^D_L(u_1,u_2)\delta(J) +
Q^C_L(u_1,u_2,J) 
\end{multline}
Last equations differs from eq. [\ref{eq:ansatz}] only in the fact that here  we do not include the contribution by the external fields and the cavity fields for the spin at the extremities of the line.
When we compute the probability distribution of the quantity 
$\mathcal H [\s_I,\s_O,\tau_I,\tau_O]$ it factorizes into the product of four terms coming from each of the lines. 
\emph{Connectivization} of the diagram, as prescribed by ref. \cite{Mlayer_method}, corresponds to removing the term  $P(u_1)P(u_2)\delta(J)$ on each line. 
Therefore, on each line we can decide if we take the contribution $Q^D$ or $Q^C$: in the first case we have a disconnected term that can be represented with a line bearing a cross, 
in the second case we have a connected term that can be represented with a line without a cross. 
In this way the diagrammatics becomes graphically equivalent to the one of the LG approach, with the addiction of extra diagmas containing cubic vertices.
We note that in this computation it is not obvious that the most divergent diagrams will be the ones with the maximal number of possible crosses, as in the standard LG expansion
that leads to DR. In fact, this is not the case for diagrams with cubic vertices. 
We obtain the following results:
\begin{itemize}
\item the connected correlation (response) function is not renormalized at one loop, since ${\cal G}^{\text{\tiny{BL}}}_R(\vec L)=0$;
\item a factor ${\cal G}^{\text{\tiny{BL}}}_C(\vec L)=\text{const}\neq0$ appears for the disconnected correlation function when $L_I=L_O=L_A=L_B=L$.
Numerically, when the four lengths are all different, the result is consistent with the behaviour ${\cal G}^{\text{\tiny{BL}}}_C(\vec L)= a+b\left(\frac{L_A}{L_B}+\frac{L_B}{L_A}\right)$ .
\end{itemize}
The detailed derivation of this result is presented in the SI, together with a numerical consistency check.

At this point one should compare this new contribution to the one obtained from the diagrams coming from the standard expansion around the LG theory,
looking at the power-law divergence when $\tau\to0$ in the limit $p\to0$.

Let's focus on $C$. 
Within the LG approach, the divergence of the diagram is of order $\tau^{-5}$.
Noticing that 
\begin{equation}
\int_1^{\infty} dL \frac{\exp\((-L(p^2+\tau)\))}{L}\,=\Gamma\[[0,(p^2+\tau)\]]
\end{equation}
and that the Incomplete Euler Gamma function behaves as $\Gamma\[[0,(p^2+\tau)\]]\simeq -\log[p^2+\tau]$ for ${(p^2+\tau)\to0}$, 
and using eqs. [\ref{eq:singlePoleRepresentation},\ref{eq:doublePoleRepresentation}], we find that
the divergence of the new diagram is at most $-\tau^{-4}\log(\tau)<\tau^{-5}$.

The new diagrams coming from cubic vertices are thus sub-dominant with respect to the standard ones in the one loop expansion of two point correlation functions.

\section*{Discussion, Conclusions and Perspectives}

In this work, we have applied the new topological expansion around the Bethe solution proposed in ref.~\cite{Mlayer_method} to the RFIM at $T=0$, numerically and semi-analytically, 
obtaining consistent results.
It is crucial that we expand around the Bethe solution because it is deeply different from the one found in a standard Landau-Ginzburg approach: 
while in the latter fluctuations do not play any role, the Bethe solution is dominated by rare fluctuations, especially at $T=0$;
this is of primary importance given that the RFIM critical behavior is controlled by a $T=0$ fixed point.

A direct consequence of the fluctuations-dominated behavior at $T=0$ is that higher order correlations do not decay faster than the average correlation, $\overline{G(x)^p}\approx\overline{G(x)}$, and this produces an effective theory with vertices of all degrees, including cubic vertices, essentially because diagrams with multiple lines between the same vertices are allowed.

We have analyzed the first two one-loop corrections to the correlation functions due to cubic vertices, finding that they give a contribution that is divergent at $D<6$, as also happens for the standard quartic diagrams. 

We also found that they give an extra contribution. However, this contribution is sub-dominant with respect to the one given by the usual one-loop diagram coming 
from the standard LG theory.
This means that, within our framework and at the 1-loop order, Dimensional Reduction is still valid at $6-\epsilon$ dimensions because the most divergent diagrams remain the super-symmetric ones 

Let us finally remark that the analyzed cubic vertices are really important already at the mean-field level (zero-th order). A peculiarity of the RFIM at $T=0$ on finite connectivity lattices is the existence of avalanches: 
this collective $T=0$ phenomenon cannot be described within the standard field theoretical treatment, while it appears naturally if $\phi^3$ vertices are introduced. 
At the critical point, the avalanches size distribution follows a power law with a nontrivial exponent $\tau$.
In our framework, we easily enough find the correct mean-field value $\tau=3/2$ that cannot be computed within the standard LG approach.
Avalanches have a fractal dimension $d_f$ that is connected to fluctuations in the $T=0$ integrated response $\chi$ via $\overline{\chi^2}/\overline{\chi}\propto L^{d_f}$.
We plan to compute the one-loop correction for $\chi^2$, i.e. for three-point functions, and obtain in this way the $\epsilon$-expansion for $d_f$.
\acknow{This research has been supported by the European Research Council under the European Unions Horizon 2020
research and innovation programme (grant No.~694925 -- Lotglassy, G Parisi) and by the Simons Foundation (grant No.~454949, G Parisi).}
\showacknow 

\clearpage


\begin{strip}
\Huge{\bf Supporting Information for\\
``New loop expansion for the Random Magnetic\\ Field Ising Ferromagnets at zero temperature''}
\end{strip}

\section{Standard diagrammatic rules for the random field Ising model}

\begin{figure}
\centering
\includegraphics[width=0.9\columnwidth]{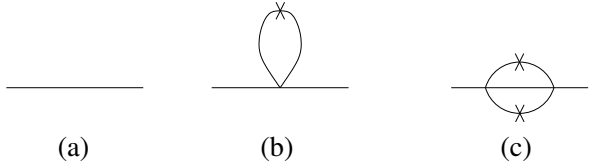}
\caption{\label{Fig:Ccon}
Most divergent diagrams for the computation of the connected correlation function $R$ in the standard theory of RFIM up to second order.}
\end{figure}

The standard way to compute the loop expansion for the Random Field Ising Model (RFIM) is to introduce an effective replicated $\phi_4$ model once the disorder has been integrated out \cite{GiardinaDeDom}.
In practice one is left with few operating rules to construct Feynmann diagrams, that we briefly recall here.

The main difference with a standard $\phi_4$ theory is that the bare propagator is composed of two parts: a connected part, that is commonly indicated with a line, going as $\widetilde{R}(p)\propto\frac{1}{p^2+\tau}$, that will contribute to the connected correlation function $\overline{\<\sigma_i \sigma_j\>_c}=\overline{\<\sigma_i \sigma_j\>-\<\sigma_i\>\<\sigma_j\>}$ and a disconnected part, indicated with a line plus a cross, $\widetilde{C}(p)\propto\frac{1}{(p^2+\tau)^2}$, that will be the dominant contribution to the disconnected correlation function 
$\overline{\<\sigma_i\>\<\sigma_j\>}$.

In practice, to build Feynmann diagrams, one should put vertices with 4 lines, that could be connected or disconnected. 
The only rule in the construction of the diagrams for the expansion of the two-point connected correlation function is that at least one connected path between the two points should be present.
Selecting only the most divergent diagrams at each order, one discovers that they correspond to the diagrams with the highest number of allowed crosses at each order in the perturbative expansion.
They are shown in Fig.~\ref{Fig:Ccon} for the connected correlation function up to second order.

\section{Solution of the Random Field Ising Model on the Bethe Lattice at $T=0$}

The solution presented in this Section is the $0^{th}$ order of the loop expansion presented in the main text. It has been presented in some detail in Refs.~\cite{Flaviano, Lucibello14}, but we find useful to reported here again for completeness.

We consider a model with Hamiltonian
$$
H = -J \sum_{(ij)\in E} \sigma_i\sigma_j - \sum_i h_i^R \s_i\;,
$$
where $J>0$ and $h_i^R$ are i.i.d.\ random variables extracted from a Gaussian probability distribution with zero mean and standard deviation $\sigma_h$. The edge set $E$ defines a Bethe lattice (BL) of finite connectivity $z$ (mathematically speaking it is a random regular graph of constant degree $z$).

Following the standard cavity method, we consider cavity fields $h_{i \to j}$  and $u_{i\to j}$ defined on each edge of the graph.
They parametrize, respectively, the marginal probability distribution on $\s_i$ in the cavity graph where edge $(ij)$ has been removed, and the marginal probability distribution on $\s_j$ in the cavity graph where all edges involving vertex $j$, but $(ij)$, have been removed.
At $T=0$ the self-consistency equations among cavity fields read
\begin{eqnarray}
h_{i\to j} &=& h_i^R + \sum_{k \in \partial i \setminus j} u_{k\to i}
\label{eq:BP1}\\
u_{i\to j} &=& \sign(h_{i\to j})\;\min(|h_{i\to j}|,J)
\label{eq:BP2}
\end{eqnarray}
where $\partial i$ is the set of neighbours of $i$, i.e.\ spins linked to $i$ via an edge of the graph.

Within the cavity method one is interested, rather than in the specific solution on a given graph, in the solution averaged over the ensemble of random graphs and random fields. To this end it is enough to solve Eqs.~[\ref{eq:BP1},\ref{eq:BP2}] in distribution sense and compute the probability distributions of cavity fields $h$ and $u$.
We call $P(u)$ the latter.

Willing to compute the correlations between spins $\sigma_0$ and $\sigma_L$ that are connected by a line of length $L$ (the path linking $\sigma_0$ and $\sigma_L$ is unique on a BL in the thermodynamic limit) we need to integrate out all the spins along the line and compute the triplet $(u_0, u_L, J_L)$, where $J_L$ is the effective coupling between $\sigma_0$ and $\sigma_L$, while $u_0$ and $u_L$ are the effective fields on $\sigma_0$ and $\sigma_L$ coming from the line.
Such a triplet can be computed in a recursive way. Let us join two chains, the first one between $\sigma_1$ and $\tau$, characterized by the triplet $(u_1,u_{\tau, 1},J_1)$, and the second one between $\tau$ and $\sigma_2$ identified by $(u_{\tau, 2},u_2,J_2)$. In order to compute the triplet describing the effective Hamiltonian between $\s_1$ and $\s_2$ we need to sum over $\tau$ and keep only the lowest energy term (we are working at $T=0$)
\begin{equation}
\begin{aligned}
\mathcal H(\s_1,\s_2) &= -\s_1u_1-\s_2u_2+\min_\f \[[-J_1\s_1 \f - h \f  -J_2 \f\s_2 \]]\\
     & \equiv E  -(u_1+u_1')\s_1 -J_{12} \s_1\s_2-(u_2+u_2')\s_2
\label{eq:H_min}
\end{aligned}
\end{equation}
with $h=h_{\tau}^R+u_{\tau,1}+u_{\tau,2}+\sum_{k\in \partial \tau \setminus 1,2}u_{k\to \tau}$, and $u_{k\to \tau}$ are independent random variables extracted from $P(u)$.
Explicit expressions for $u_1'$, $u_2'$ and $J_{12}$, assuming $J_2\geq J_1\geq 0$, are given in Table~\ref{table:BPrules}, where $h_{\pm} = \big(h \pm \sign(h) (J_2-J_1)\big)/2$.
We went from the two initial triplets $(u_1,u_{\tau, 1},J_1)$, and $(u_{\tau, 2},u_2,J_2)$ to the new one $(u_1+u_1',u_2+u_2', J_{12})$, with the insertion of $z-2$ cavity fields 
acting on the central spin $\tau$.

\begin{table}[ht]
\centering
\caption{\label{table:BPrules}
Rules for evolving cavity fields and effective coupling in the computation of correlations at $T=0$.}
\begin{tabular}{c|c c c c}
  &  $u_1'$ & $u_2'$ & $J_{12}$  \\
 \hline
$|h|>J_2+J_1$ & $\sign(h) J_1$ & $\sign(h) J_2$ & 0 \\
$J_2-J_1<|h|<J_2+J_1$ & $h_-$ & $h_+$ & $\frac{J_1+J_2-|h|}{2}$ \\
$|h|<J_2-J_1$ & 0 & $h$ & $J_1$ \\
\end{tabular}
\end{table}

In practice we start from a population $P_{L=1}(u_0,u_1,J_1)$ of triplets all equal to $(0,0,J)$.
To evolve the population $P_{L-1}$ into population $P_{L}$ we follow the rules summarized in Table~\ref{table:BPrules}, where each triplet of the population $P_{L-1}$ is joined to a triplet $(0,0,J)$ and $z-2$ cavity fields $u_{k\to \tau}$ extracted from $P(u)$ are added on the central spin.

Unfortunately this procedure is very ineffective, because at each step a constant fraction of the population (the one satisfying the condition $|h|>J_{L-1}+J$) 
produces a  new triplet with $J_L=0$. Given that $J_L=0$ is a fixed point of the iteration, 
the part of the population keeping information about branches with non-zero effective couplings shrinks exponentially fast during the iteration.

\begin{figure}
\centering
\includegraphics[width=0.9\columnwidth]{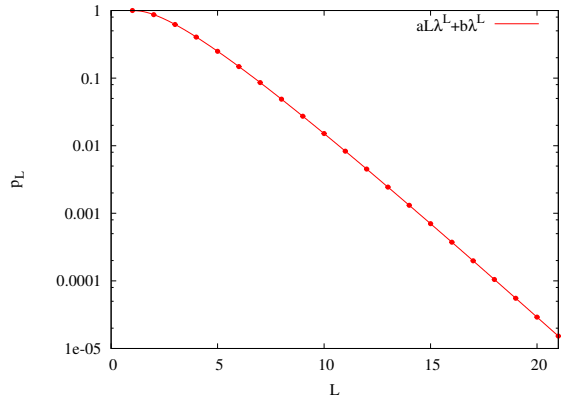}
\caption{\label{Fig:numerical_pL}
Exponential decay of the probability to have $J_L\neq0$ on a chain of length $L$ in a BL, data computed at the critical field $\sigma_{h,c}= 1.037$ for $z= 3$. Errors are smaller than points.}
\end{figure}

\begin{figure}
\centering
\includegraphics[width=0.9\columnwidth]{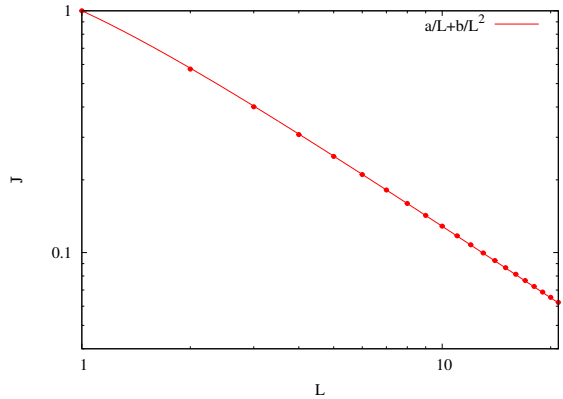}
\caption{\label{Fig:numericalJave}
Average coupling computed on the population of triplets $(u_0,u_L,J_L)$ with $J_L\neq0$. Its decay follows the law $J_L=\frac{a}{L}+\frac{b}{L^2}+o(L^{-2})$. Data are computed at the critical field $\sigma_{h,c}= 1.037$ for $z= 3$. Errors are smaller than points.}
\end{figure}

To amplify this signal, we evolve two populations of the same size: one population keeps the pairs $(u_0,u_L)$ along branches with $J_L=0$, 
while the second one stores the triplets along branches with $J_L\neq0$. At the same time we measure the probability $p_L=\mathbb{P}[J_L\neq0]$, 
that is the relative weight of the second population to the first one, which is found to decay exponentially fast with $L$: 
$p_L=a L \lambda^L+b\lambda^L+o(\lambda^L)$ as shown in Fig. \ref{Fig:numerical_pL}.
$\lambda$ is the largest eigenvalue associated to the linearization of the BP eqs.~[\ref{eq:BP1},\ref{eq:BP2}] around the fixed point.
At the critical point, $\s_{h}=\s_{h,c}$, $\lambda(\s_{h,c})=1/(z-1)$ holds.
The average coupling on the second population decays as $\overline{J_L}=\frac{a}{L}+\frac{b}{L^2}+o(L^{-2})$ as shown in Fig. \ref{Fig:numericalJave}.
We see that $\overline{u_0 u_L}\propto \frac{1}{L^2}$ on the population with $J\neq0$ while $\overline{u_0 u_L}\propto L\lambda^L$ on the population with $J=0$.
Moreover $\overline{u_0^2 J}-\overline{u_0^2}\,\overline{J}\propto \frac{1}{L^2}$ on the population with $J\neq0$.

Once we have $p_L$ and the two populations at each length $L$, it is quite simple to compute correlation functions.
Indeed, given a triplet $(u_1,u_2,J_{12})$ associated to a path, where the internal spins have been integrated out, the effective two spin Hamiltonian reads
\begin{equation}
\mathcal H [\s_1,\s_2] = -h_1 \s_1  -J_{12} \s_1\s_2 -h_2 \s_2.
\end{equation}
with $h_1=h^R_1+u_1+\sum_{k1=1}^{z-1}u_{k1}$, $h_2=h^R_2+u_2+\sum_{k2=1}^{z-1}u_{k2}$, and $u_{k1}$, $u_{k2}$ extracted from $P(u)$.
At zero temperature the Gibbs measure is concentrated on the ground state $(\s^*_1,\s^*_2)$ of the Hamiltonian, that can be easily computed using	 the rules listed in Table~\ref{table:GSrules}.

\begin{table}[ht]
\centering
\caption{\label{table:GSrules}
Rules for computing the ground state configuration given the triplet of cavity messages $(h_1,h_2,J_{12})$.}
\begin{tabular}{c|c}
  & $\s^*_1$, $\s^*_2$  \\
 \hline
$|h_1| < \min(J_{12},|h_2|)$  &  $\s^*_1=\s^*_2 =\sign(h_2)$\\
$|h_2| < \min(J_{12},|h_1|)$  &  $\s^*_1=\s^*_2 =\sign(h_1)$\\
$J_{12} < \min(|h_1|,|h_2|)$  &  $\s^*_1=\sign(h_1)$, $\s^*_2=\sign(h_2)$\\
\end{tabular}
\end{table}

Since we are at $T=0$, the disconnected correlation function is given by
\begin{equation}
C_{ij} \equiv \overline{\langle\s_i \rangle \langle\s_j\rangle} \equiv \overline{\s_i^*\s_j^*}
\end{equation}
where $\s^*$ is the ground state configuration, computed according to the rules listed in Table~\ref{table:GSrules}. The connected correlation function $C^{con}_{i j} = \overline{\langle\s_i \s_j\rangle_c}$ is ill defined since it is identically equal to zero at $T=0$, therefore we work with the response
\begin{equation}
R_{ij} = \frac{1}{2}\ \overline{1 - \s^*_i\,\langle\s_i\rangle_j}
\label{eq:Rij}
\end{equation}
where $\langle \cdot\rangle_j$ denotes the expectation over the ground state of the system conditioned to the flipping of the spin $\s_j$ , i.e. $\langle\s_j\rangle_j = -\s^*_j$. 
This can be achieved adding a field $h_j=-\s^*_j \cdot \infty$  on the spin $\s_j$. 
An alternative and more general definition of response would be $R_{ij} = \frac{1}{2}\ \overline{\s^*_j(\s^*_i - \,\langle\s_i\rangle_j)}$.
Since in the RFIM the couplings are ferromagnetic, the spin $\s_i$ can only flip in the same direction of the flip of $\s_j$, therefore the two definitions are equivalent and $0\leq R_{ij}\leq 1$.
It can be shown that the response can be expressed as the zero temperature limit of an opportunely normalized connected correlation function, that is
\begin{equation}
R_{ij}=\lim_{\beta\to\infty} \overline{\left[\frac{\langle \s_i \s_j\rangle_c}{1-\langle \s_i\rangle^2}\right]}.
\end{equation}

\begin{figure}
\centering
\includegraphics[width=0.9\columnwidth]{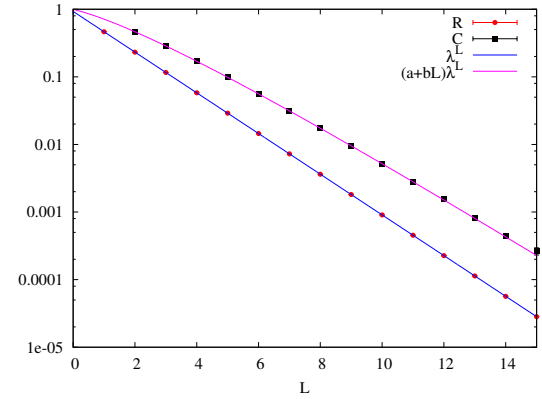}
\caption{\label{Fig:corr_line}
Connected and disconnected correlation functions as a function of the distance $L$ on the Bethe lattice. Data are computed at the critical field $\sigma_{h,c}= 1.037$ for $z= 3$.}
\end{figure}

In terms of the probability law of random triplets $(h_1,h_2,J_{12})$ the correlation functions can be written as
\begin{eqnarray}
C_{12}&=&\mathbb{P}\left[J_{12}>\min(|h_1|,|h_2|)\right]+\nonumber\\
&&\mathbb{E}\left[\sign(h_1)\sign(h_2)\, |\, J_{12}<\min(|h_1|,|h_2|)\right]
\label{c-twospins}\\
R_{12}&=&\mathbb{P}\left[J_{12} > |h_1|\right]=\mathbb{P}\left[J_{12} > |h_2|\right].
\label{r-twospins}
\end{eqnarray}
On chains, $h_1$ and $h_2$ are positively correlated,therefore $C_L \ge R_L$.  Notice that only events with a non-zero effective couplings contribute to the response function: this is the reason why amplifying the population of cavity messages with $J_L\neq 0$ is mandatory to have a precise measurement of correlations in the $T=0$ limit.

In Fig.~\ref{Fig:corr_line} we show the connected and disconnected correlation functions at distance $L$, averaged over the population of the triplets generated as explained before, in a BL with fixed connectivity $z=3$, at zero temperature and critical standard deviation $\s_{h,c}=1.037$ for the external field.
We find the the connected correlation function decays as $R_L \propto \lambda^L$, with $\lambda=\frac{1}{z-1}$, while the disconnected correlation function is larger and decays as $C_L = (a L + b) \lambda^L$, as already found analytically in Refs.~\cite{Flaviano, Lucibello14}.  

The corresponding susceptibilities can be computed by summing over all the vertices of the graph
\begin{equation}
\chi_{disc}=\sum_j C_{0j}=\sum_L n_L C_L 
\label{eq:chi0}
\end{equation}
where $n_L$ is the number of spins at distance $L$ that in a BL is $n_L=\frac{z}{z-1}\cdot (z-1)^{L}$. 
Substituting $n_L$ and $C_L$ in the equation for $\chi$ one gets:
\begin{align*}
\chi_{disc}=&\frac{z}{z-1}\sum_L \((\lambda(z-1)\))^L \cdot (a L+b)= \\
=&\frac{z}{z-1}\bigg[\frac{a (z-1) \lambda}{\((1-(z-1)\lambda\))^2}+\frac{b}{1-(z-1)\lambda}\bigg]
\end{align*}
At the critical point $\lambda(\s_{h,c})=\frac{1}{z-1}$ and the susceptibility diverges. 
How can we relate this computation of the susceptibility on the Bethe lattice to the perturbative expansion for a finite dimensional model in dimension $D$ around the Bethe theory?
Following ref.  \cite{Mlayer_method}, the zeroth order expansion for the susceptibility is just eq. [\ref{eq:chi0}] where $n_L$ is replaced with the number of non-backtracking paths of length $L$ 
starting from a point in a $D$ dimensional lattice:
$n_L\propto (2D-1)^L$. Therefore at zeroth order the expansion predicts a divergence located at the same critical point $\s_{h,c}$ of a Bethe lattice with connectivity $z=2D$.

\subsection{One loop BL correction}

\begin{figure}
\centering
\includegraphics[width=0.9\columnwidth]{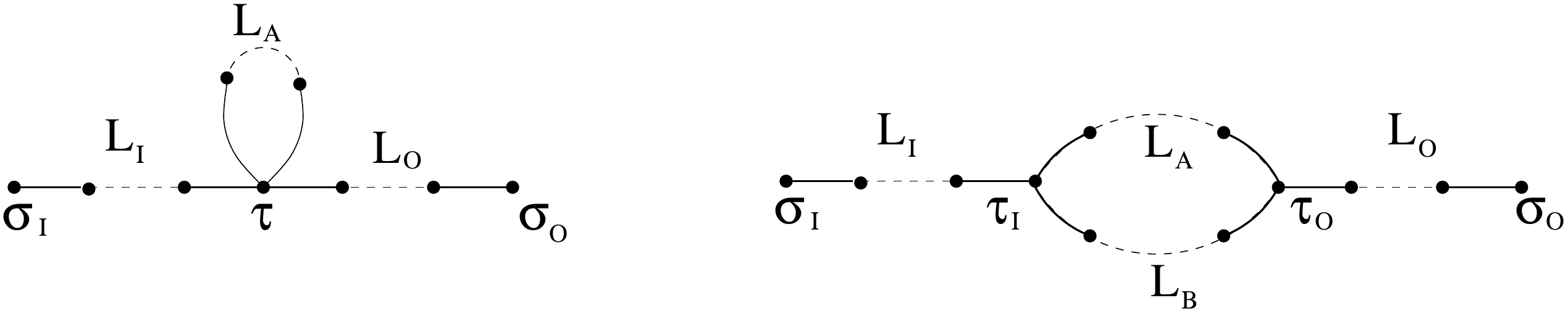}
\caption{\label{Fig:loop_diagram}
One loop topological diagrams important for the first order expansion around the Bethe lattice: they can have vertices with four lines (left), or vertices with three lines (right).
The right one is not present in the standard LG theory but is present in the BL expansion.}
\end{figure}

The first order in the BL expansion considers the presence of structures with one spatial loop, as reported in Fig. \ref{Fig:loop_diagram}. 
In this section, we compute the one-loop correction to the connected and disconnected correlation function, coming from the topological structure in the right part of Fig.~\ref{Fig:loop_diagram},
that is the one that gives an additional term with respect to the standard Landau-Ginsburg (LG) expansion. 
Following the prescription of ref. \cite{Mlayer_method}, we should compute the correlation on a BL in which such structure has been manually injected 
and subtracting the values of the correlation computed on the two paths $L_I+L_A+L_O$, $L_I+L_B+L_O$, supposed as independent.

Operatively we build the loop putting together four triplets or couples extracted independently from the populations of single Bethe lines obtained as explained in the previous section: 
two of length $L_I$ and $L_O$ for the external lines, and two of length $L_A$ and $L_B$
for the internal lines of the loop. 
The internal lines of the loop, with triplets $(u_{\tau_I}^A,u_{\tau_O}^A, J_A)$ and $(u_{\tau_I}^B,u_{\tau_O}^B, J_B)$, 
will just result in a new triplet whose coupling is the sum of the couplings: $J_T=J_A+J_B$
and whose fields are the sum of the fields $u_{\tau_I}^T=u_{\tau_I}^A+u_{\tau_I}^B$. 
Then the new triplet is attached to the external legs, as in eq. (\ref{eq:H_min}), with the only difference being that in $\tau_I$ and $\tau_O$ there are just $z-3$ additionally 
cavity fields (instead of $z-2$ ones) extracted from $P(u)$.
We then end up with a new triplet describing the loop. We compute the correlations implied by this triplet and subtract the correlation implied by the paths $L_I+L_A+L_O$ and $L_I+L_B+L_O$ 
considered as independent: This is the one-loop contribution to the correlation function, that we will indicate with $G_{loop}(\vec L)=G_{loop}({L_I,L_O,L_A,L_B})$.
In the following we report the one-loop results for $G=C,R$, that we obtain looking numerically to the behaviors when $L_A$ or $L_B$ are small or large.

\begin{figure}[t]
\begin{center}
 \includegraphics[width=0.9\columnwidth]{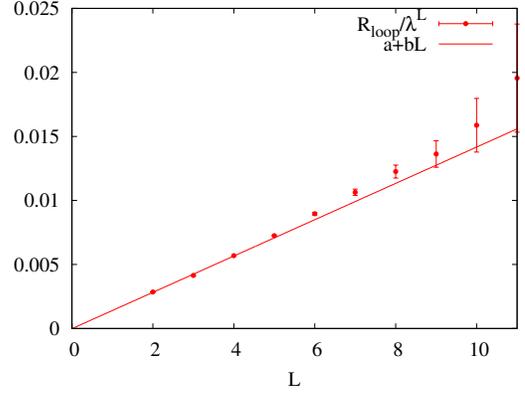}
 \vspace{1cm}
\caption{\label{Fig:conn_tadpole}
Absolute value of the one-loop BL contribution to the connected correlation function divided by $\lambda^L$,
when $L_I=L_B=L_O=3$ and $L_A=L$, as function of distance $L$. The behaviour $R_{loop}(3,3,L,3)\propto \lambda^L L$ is evident. The sign of the contribution is negative. Data is computed at the critical field $\sigma_{h,c}= 1.037$ for $z= 3$.}
\end{center}
\end{figure}

Let us first analyze the response.
We know that the contribution to $R$ given by the external legs should, in any case, be proportional to $\lambda^{(L_I+L_O)}$, 
because the diagram should be connected to contribute to the connected
correlation function.
Thus we concentrate on the internal legs.
First of all, we fix also $L_B$ to a finite value $L_B=3$ and we measure the contribution of the loop to the response function as a function of $L_A=L$.
We measure the behavior 
\begin{equation}
 R_{loop}({L_I,L_O,L,3})\propto L\lambda^{L+L_I+L_O},
 \label{eq:tadpoleR}
\end{equation}
as shown in Fig. \ref{Fig:conn_tadpole}. As expected, the behaviour is the same of the one-loop diagram coming from the standard theory, that is the left diagram of Fig.~\ref{Fig:loop_diagram}.
In fact, two $\phi_3$ vertices reduces to a tadpole $\phi_4$ vertex once one internal line is fixed to a finite length. For a tadpole $\phi_4$ diagram,
LG theory predicts that the maximal divergent contribution comes from the second diagram of Fig. \ref{Fig:Ccon}, that has indeed the same behavior of as in eq. [\ref{eq:tadpoleR}].

\begin{figure}[t]
\begin{center}
 \includegraphics[width=0.9\columnwidth]{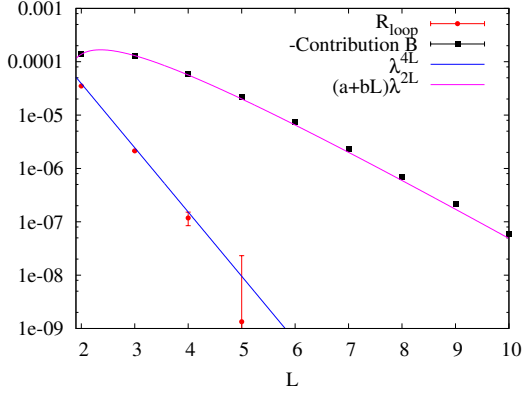}
 \vspace{1cm}
\caption{\label{Fig:oneloop_conn_rfim}
One-loop BL contribution to the connected correlation function
when $L_I=L_O=3$ and $L_A=L_B=L$, as function of the length of internal lines $L$. 
The total contribution is the sum of Contribution A and B. They separately go like $(a+bL)\lambda^{2L}$, but they have opposite sign.
Once they are summed, the result decays as $\lambda^{4L}$. Data are computed at the critical field $\sigma_{h,c}= 1.037$ for $z= 3$.}
\end{center}
\end{figure}

Next, we fix $L_A=L_B$ and we measure $R_{loop}(L_I,L_O,L,L)$.
It receives contributions from two different diagrams: we call ``Contribution A'' the one coming from the loop with both $J_A$ and $J_B$ different from zero, 
that is, in the language of eq. [36] of the main text, taking the contribution coming from $Q^C_L(u_1,u_2,J)$ on both lines;
 ``Contribution B'' instead, is the one from a loop with just one coupling different from zero, 
that is, taking a contribution coming from $Q^C_L(u_1,u_2,J)$ on one line, and a contribution coming from $Q^D_L(u_1,u_2)\delta(J)$ on the other line.
(Please note that it is not possible to take contribution from $Q^D_L(u_1,u_2)\delta(J)$ on both internal lines because it will result in a disconnected loop that 
gives zero contribution to connected correlation functions).
Separately, contribution A and B, multiplied by their occurrence probabilities, have a dominant behaviour in $L$ of the type $(a+bL)\lambda^{L_I+L_O+2L}$, but they have opposite sign. 
When summing the two contributions, the dominant term in $L$ is exactly cancelled, and the total contribution is left with the subdominant part
\begin{equation}
 R_{loop}({L_I,L_O,L,L})\propto \lambda^{L_I+L_O}\lambda^{4L},
 \label{eq:loopR}
\end{equation}
as shown in fig. \ref{Fig:oneloop_conn_rfim}. 
As described in the main text, we can write $R_{loop}(\vec L)=\lambda^{\Sigma(\vec L)}{\cal G}_R(\vec L)+o(\lambda^{\Sigma(\vec L)})$, with $\Sigma(\vec L)=L_I+L_O+L_A+L_B$. Eq. [\ref{eq:loopR}] corresponds to ${\cal G}_R(\vec L)=0$.

Things are different for the disconnected correlation function. Also in this case, first of all we fix $L_B$ to a finite value and we 
measure the behaviour of the loop as a function of $L_A=L$. We measure the behavior 
\begin{equation}
 C_{loop}({L_I,L_O,L,3})\propto L\lambda^{L+L_I+L_O},
 \label{eq:tadpoleC}
\end{equation}
that again is the same contribution of the tadpole diagram from the usual $\phi_4$ LG theory. 
Then we put $L_A=L_B=L_I=L_O=L$, and we observe
\begin{equation}
 C_{loop}({L,L,L,L})\propto \lambda^{4L},
 \label{eq:loopC}
\end{equation}
that corresponds to ${\cal G}_C(L,L,L,L)=const$.

\begin{figure}[t]
\begin{center}
 \includegraphics[width=0.9\columnwidth]{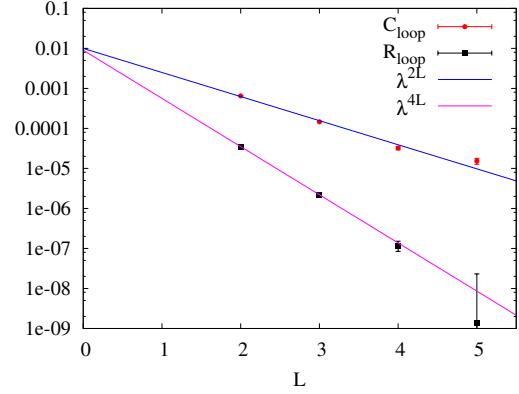}
 \vspace{1cm}
\caption{\label{Fig:oneloop_conn_disc_rfim}
Comparison between the one-loop BL contribution to the connected and disconnected correlation function
when $L_I=L_O=3$ and $L_A=L_B=L$, as function of the length of internal lines $L$. Data are computed at the critical field $\sigma_{h,c}= 1.037$ for $z= 3$.}
\end{center}
\end{figure}

We now want to compute the one-loop correction to the susceptibilities. In an analogous way to eq. [\ref{eq:chi0}], 
we should account for all the subgraphs of the type of the right dyagram in Fig. \ref{Fig:loop_diagram}
that are presents in a finite-dimensional lattice.
The computation can be done exactly using the number of non-backtracking paths, however, the large $L$ behavior of this counting factor is also captured if we assume that the number of paths from 0 to $x$ of length $L$ is
given by the random walk probability of  reaching $x$ in time $L$  in $D$ dimensions
multiplied by the number of generic non-backtracking paths of length $L$ starting from 0: 
$n_L(x)\propto \frac{(z-1)^L}{L^D} e^{-x^2/(2L)}$.
In the same way we can compute $n_{L_A,L_B}(x,y)$, defined as the number of paths of length $L_A$ and $L_B$ that have the same starting and ending point $x$ and $y$:
$n_L(x,y)\propto\frac{(z-1)^{L_A+L_B}}{(L_AL_B)^D} \left(e^{-\frac{(x-y)^2}{2}(\frac{1}{L_A}+\frac{1}{L_B})}\right)$.
The one loop correction to the susceptibility associated to a generic correlation function $G$ is thus:
\begin{align*}
 \chi_{loop}\propto &\sum_{L_I,L_A,L_B,L_O} \sum_{x,y,d}n_{L_I}(x)n_{L_A,L_B}(y-x)\cdot\\
 &\cdot n_{L_O}(d-y)G_{loop}(\vec L).
\end{align*}
Replacing the sums over $x,y,d$ with integrals, and performing the integrals we obtain:
\begin{align*}
 \chi_{loop}\propto& \sum_{\vec L} G_{loop}(\vec L)(z-1)^{\Sigma(\vec L)}\frac{1}{\((L_A+L_B\))^{D/2}}.
\end{align*}
It is now clear that the contribution to the connected susceptibility is always finite at the critical point, because $R_{loop}(\vec L)$ decays more rapidly than $(z-1)^{-\Sigma(\vec L)}$.
Things are different for the disconnected susceptibility.
In this case, we can apply the Ginzburg criterion to identify the upper critical dimension: we look at $\frac{d\chi}{d \lambda}$, that is divergent at the critical point in $D\le6$.
At this point, one could think that Dimensional Reduction is broken at $D=6$. In fact for $D<6$, we have cubic diagrams, not of the type of the super-symmetric ones, that are important.
However, as we explained in the main text, once we compare their divergence with the divergence of the standard one-loop $\phi_4$ diagrams,
we discover that their contribution is sub-dominant with respect to the usual $\phi_4$ term. This implies that DR is still valid at $6-\epsilon$ dimensions.

To conclude, we just mention that until now we do not know the exact behavior of $C_{loop}$ as a function of the length of the legs. 
Having measured $C_{loop}(L,L,L,L)\propto \lambda^{4L}$ 
we can think to different cases:
\begin{itemize}
 \item A) $C_{loop}(\vec L)\propto\lambda^{L_I+L_A+L_B+L_O}$
 \item B) $C_{loop}(\vec L)\propto\lambda^{L_I+L_A+L_B+L_O}(\frac{L_A}{L_B}+\frac{L_B}{L_A})$
 \item C) $C_{loop}(\vec L)\propto\lambda^{L_I+L_A+L_B+L_O}(\frac{L_I}{L_B}+\frac{L_I}{L_A}+\frac{L_O}{L_B}+\frac{L_O}{L_A})$ 
\end{itemize}
We expect that the presence of terms of the type $\lambda^L/L$ should signal the presence of squared disconnected correlation function 
\footnote{We somehow expect the presence of important square correlations at zero temperature, see the Conclusions}. In fact we measured numerically that
$\overline{(u_0 u_L)^2}\propto\lambda^L/L$, and we expect that $\overline{(\<\sigma_0\>\<\sigma_L\>)^2}\propto\overline{(u_0 u_L)^2}$, 
given that $\overline{\<\sigma_0\>\<\sigma_L\>}\propto\overline{u_0 u_L}$.
To understand which terms are present, we measure $C_0\equiv C_{loop}(L_I,L_O,L,L)$ and $C_1\equiv C_{loop}(L_I,L_O,2L,L)$, at fixed, 
finite values of $L_I$ and $L_O$, and we look at them as a function of $L$.
We obtain the behaviours: $C_0\propto\lambda^{2L}$, $C_1\propto\lambda^{3L}$. The ratio $Q(L_I,L_O)\equiv\frac{C_0/ \lambda^{2L}}{C_1 /\lambda^{3L}}$ is independent from $L_I$ and $L_O$.
This result tells us that the case C) is not present. Indeed this is what we expected: the case C) corresponds to a connected loop, but we know that the connected correlation, 
that can receive contribution only by a connected loop, is not renormalized at one loop. We thus expect that the connected loop gives no contribution to $C$, as found.
We numerically find that $Q=0.96$: if the situation A) were the only present, $Q=1$, while in the case B) $Q=0.8$:
to recover the measured $Q=0.96$ we need a linear combination of the two cases.
From the numerical computation, we thus expect the one-loop contribution to the disconnected correlation function to have the form 
$$C_{loop}(\vec L)\propto\lambda^{\Sigma(\vec L)}\[[a+b\((\frac{L_A}{L_B}+\frac{L_B}{L_A}\))\]],$$
with $a=1$, $b=0.1$.

\section{BL results for the distribution of Avalanches}

The distribution of the size of the avalanches $s$ at the critical point is expected to be
\be
P(s)=\frac{1}{s^{\rho}}
\label{eq:P(chi)}
\ee
with $\rho$ the critical exponent for the avalanches whose value on the BL is $\rho_{MF}=\frac{3}{2}$.
This distribution can be obtained in the framework of percolation on the BL (see \cite{GarelMonthus} and refs. therein).

\begin{figure}
\centering
\includegraphics[width=0.7\columnwidth]{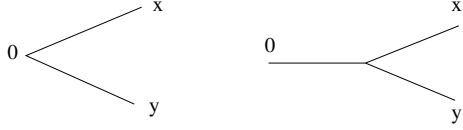}
\caption{\label{Fig:vertice_phi_3_chi_2}
Important topological diagrams for the calculation of the leading term for $\chi^2$ in the standard $\phi_4$ theory (left) and in a theory with $\phi_3$ vertices (right).}
\end{figure}

We explained in the main text that $s$ is proportional to the susceptibility associated to the connected correlation function: $\chi=\sum_x \<\sigma_0\sigma_x\>_c$.
Given that $\overline{\chi}\propto\tau^{-1}$, the distribution [\ref{eq:P(chi)}] in the MF region
implies that $\overline{\chi^2}\propto\tau^{-3}\neq \overline{\chi}^2$.
This result cannot be recovered from the LG theory. In this case, in fact, the global magnetization is the only important variable, 
there are no fluctuations in the magnetization nor in the susceptibility, for which therefore we can write $\overline{\chi^2}=\overline{\chi}^2\propto\tau^{-2}$. 
Let us now look in detail to what are the field-theoretical predictions on
$\chi^2=\sum_{x,y} \<\sigma_0\sigma_x\>_c\<\sigma_0\sigma_y\>_c$, for which we have to look to three point functions.
If we admit that there are only $\phi_4$ vertices, as in the MF FC model, the diagram with no loop is the left one in Fig. \ref{Fig:vertice_phi_3_chi_2}. 
Giving that each line corresponds to a connected propagator and thus bring a factor $\tau^{-1}$,
the left diagram will be associated at the critical point to a divergence of the type $\overline{\chi^2}\propto\tau^{-2}$, recovering the FC MF result.
If now we imagine that the associated field theory includes also $\phi_3$ vertices, the situation will change: the right diagram in Fig. \ref{Fig:vertice_phi_3_chi_2} is possible,
leading to a critical behaviour: $\overline{\chi^2}\propto\tau^{-3}$.

We have announced that, following ref. \cite{Mlayer_method}, diagrams with $\phi_3$ vertices should be present in the field-theoretical description of the RFIM at $T=0$ when expanding
around the finite connectivity Bethe solution: 
in this section, we have shown that their presence is perfectly compatible with the MF description of the avalanches, for which we can recover the critical exponent $\rho^{MF}=\frac{3}{2}$,
in contrast to the standard FC $\phi_4$ theory that cannot justify 
the probability distribution of the avalanches. 

In ref. \cite{TarjusAvalanches}, the following connection between avalanches and DR is stated:
DR breaks down due to avalanches if they are ``big enough'', more precisely if the fractal dimension $d_f$ of the largest typical critical avalanches satisfies the condition
$d_f=D-d_{\phi}$, with $D$ the spatial dimension and $d_{\phi}$ the scaling dimension of the field near the relevant zero-temperature fixed point.
In a way analogous to Ref.~\cite{TarjusAvalanches}, we have seen that the $\phi_3$ diagrams will not automatically destroy DR: in particular, at one loop they are sub-dominant with respect to the standard $\phi_4$ ones implying that at $D=6-\epsilon$, DR is preserved. 

\section{Ansatz for coupling and fields at distance L}

In this section, we verify our previous numerical results using a different method. 
We introduce an Ansatz $P_L(u_0,u_L,J)$ for the joint distribution of the effective coupling and fields between two spins at distance $L$ in a BL, which should capture the leading behviour at large $L$. 
We assume the form
\begin{equation}
\begin{aligned}
P_L(u_0,u_L,J)=& \delta(J)\bigg[ P(u_0)P(u_L)- c_0 L \lambda^L g(u_0)g(u_L)\\
& - c_1 L \lambda^L g'(u_0)g'(u_L)- c_2 L \lambda^L g''(u_0)g''(u_L) \bigg]\\
&+ a L^2 \lambda^L \rho\, e^{-\rho J L}g(u_0)g(u_L) +o(L\lambda^L)
\end{aligned}
\label{eq:PLinit}
\end{equation}
and check it's consistency.
$P(u)$ is the already mentioned Bethe distribution of cavity fields, while $g(u)$ is the eigenfunction associated to the largest eigenvalue $\lambda$ with respect to a perturbation of $P(u)$ \cite{ParisiRicciRizzo}.
$g(u)$ is symmetric, therefore $g'(u)$ is anti-symmetric. We impose $\int_{-\infty}^{\infty} g(u)du = 1$.

We impose normalization:
\begin{equation}
1=\int \text{d}u_0\text{d}u_L\text{d}J\ P_L(u_0,u_L,J) = 1- c_0 L\lambda^L  + a L \lambda^L,
\end{equation}
obtaining the relation $c_0= a $.

The functional form \ref{eq:PLinit} has to reproduce itself when attaching  two chains  to create a new one of length $L_1+L_2=L$, according do the rules of Table~\ref{table:BPrules}. Using the symbol $\otimes$ to denote the iteration in distribution of two chains according to these rules, or the addition of a field to an extremity of a chain, we have to check that
\begin{equation}
P_L = P_{L_{1}}\otimes Q_{z-2} \otimes P_{L_{2}}
\end{equation}
where $Q_{z-2}$ is the distribution of the sum of $z-2$ cavity fields extracted from $P(u)$ plus the random external field. More explicitely, we have
\begin{equation}
\begin{aligned}
P_L(u_0,u_L,J)
=&\int dP_{L_{1}}(v_0,v_{\tau,1},J_1) \,dQ_{z-2}(h)\,dP_{L_{2}}(v_{\tau,2},v_L,J_2)\\
\times&\delta(u_0 - v_0 - f_1(v_{\tau,1}+v_{\tau,2}+h,J_1,J_2))\\
\times&\delta(u_L - v_L - f_2(v_{\tau,1}+v_{\tau,2}+h,J_1,J_2))\\
\times&\delta(J - f_J(v_{\tau,1} + v_{\tau,2} + h,J_1 , J_2)),
\label{eq:selfcons}
\end{aligned}
\end{equation}
where the functions $f_1$, $f_2$ and $f_3$ 
can be deduced from Table \ref{table:BPrules}.
A careful computation of the leading order terms in the right hand side, 
shows that the Ansatz \eqref{eq:PLinit} holds, provided that
\begin{equation}
a = \frac{\rho}{2 \hat{P}(0)} , \quad c_2=0,
\end{equation} 
where 
\begin{equation}
\hat P(h) =\int \text d u\,\text d v\,\text d h \ g(u)g(v)Q_{z-2}(h) \ \delta(h-(u+v+h)) 
\end{equation}
It turns out that $\rho$ and $c_1$ are left undetermined. 
The final form for the Ansatz is thus given by:
\begin{equation}
\begin{aligned}
P_L(u_0,u_L,J)=& \delta(J)\bigg[ P(u_0)P(u_L)- a\,L \lambda^L g(u_0)g(u_L)+ \\
 & - c_1 L \lambda^L g'(u_0)g'(u_L) \bigg]+\\
&+ a L^2 \lambda^L \rho\, e^{-\rho J L}g(u_0)g(u_L) 
\end{aligned}
\label{PLfinal}
\end{equation}
In this form, the Ansatz is normalized and stable under the merging of two chains up to $o(L\lambda^L)$ terms. 
The coefficient $\rho$ can be derived using a few additional arguments, see next Section.
This form for the Ansatz is compatible with what we know from the previous sections: the coupling $J$ either is exactly 0 or, with
a probability of order $L\lambda^L$ , is of order $1/L$.
The two effective fields and the coupling are independently distributed when conditioning on the event $J_L>0$. The two fields have correlation of order $O(L\lambda^L)$ when conditioning on  $J_L=0$ instead.
Moreover, the Ansatz reproduces the numerical behavior of the correlation functions at length $L$.

Now we want to reproduce the numerical results for the loop contribution to the correlation functions using the Ansatz.
In order to obtain the joint distribution $P_{\text{loop}}(u_{I},u_{O}, J_{\text{loop}})$ of the effective fields and coupling for two spin at the extremities of a loop as in the right part of Fig. \ref{Fig:loop_diagram},
we convolve the two internal branches $L_A$ and $L_B$, yielding a distribution that we call $(P_{L_A}* P_{L_B})$ on the two internal spins,  and attaching the external legs $L_I$ and $L_O$:  
\begin{multline}
P_{\text{loop}}=
P_{L_{I}}\otimes Q_{z-3}\otimes \left(P_{L_{A}}*P_{L_{B}}\right)\otimes Q_{z-3}\otimes P_{L_{O}}
\label{eq:Ploop}
\end{multline}

We have already said that the loop contribution to the observable is given by the value of the observable computed on the loop minus the observable computed on the two paths $L_I+L_A+L_O$ and
$L_I+L_B+L_O$ considered as independent. We can easily obtain this loop correction defining the ``topologically connected'' loop distribution $\tilde{P}_{\text{loop}}(u_I,u_O,J_{\text{loop}})$
as in eq. (\ref{eq:Ploop}) but substituting to $P_L$ the (improper) distribution $\widetilde{P}_L$ given by\begin{equation}
\begin{aligned}
\widetilde{P}_L(u_0,u_L,J)=& \delta(J)\bigg[- a\,L \lambda^L g(u_0)g(u_L)+ \\
 & - c_1 L \lambda^L g'(u_0)g'(u_L) \bigg]+\\
&+ a L^2 \lambda^L \rho\, e^{-\rho J L}g(u_0)g(u_L),
\end{aligned}
\end{equation}
that is, the same as $P_L$ but without its asymptotic term.
In this way, the loop correction is just the mean value of the observable on $\tilde{P}_{\text{loop}}(u_I,u_O,J_{\text{loop}})$.
The loop correction for both the connected and disconnected correlation function computed on $\tilde{P}_{\text{loop}}(u_I,u_O,J_{\text{loop}})$ gives 0.
While this is in agreement with the numerical computation for $R$, we had been expecting a non-zero contribution for $C$.
However, being the Ansatz consistent up to order $O(L\lambda^L)$, it could only give a contribution to $C_{loop}(L,L,L,L)=O(L\lambda^{\Sigma(\vec L)})$ that in fact is not present
from the numerical analysis (Please notice that higher contributions are prohibited for symmetry reasons).
Thus the analytical Ansatz predictions are fully compatible with the numerical results up to the chosen order.

To go to next order, we should introduce terms $O(\lambda^L)$ in the ansatz. Unfortunately, the
addition of new terms in the original Ansatz makes the computation much more involved, and we did not perform it entirely.
In particular these new terms should take into account correlations between fields and coupling in the $J\neq0$ part, as found from the numerical analysis.
However terms in the $J=0$ part can be added without much effort, in particular we added the terms $-b_0\lambda^L g(u_0)g(u_L)-b_1\lambda^L g'(u_0)g'(u_L)-b_2\lambda^L g''(u_0)g''(u_L)$ and checked how they behave under iteration.
Imposing normalization and self-consistency we do find that $b_0=b_2=0$ and $b_1=\frac{1}{(2\hat{P}(h=0))^2}$.
The addition of this new term gives no contribution to $R$ while it gives a contribution $C_{loop}(L,L,L,L)=O(\lambda^{\Sigma(\vec L)})$ for the disconnected correlation function, 
as found from the numerical computation. We stress however that we lack some terms coming from the correction of the $J\neq0$ part of the Ansatz to order $O(\lambda^L)$ that do not allow
us to compute exactly $C_{loop}(L,L,L,L)$ at order $O(\lambda^{\Sigma(\vec L)})$.

\section{Computation of the mean coupling decay}

The analytical Ansatz presented in the previous Section requires the knowledge of 2 parameters: $c_1$ and $\rho$. Here we show how to compute the latter in a very effective way. We follow the ideas of Ref.~\cite{Flaviano}, but correcting an error made in that work.

In practice we are interested in computing the mean value of the effective coupling at distance $L$ along the branches of the BL, where the coupling is non zero
\begin{equation}
\langle J \rangle_{J>0} = \frac{1}{\rho L}
\end{equation}
Without loss of generality and to make analytical expressions more compact we fix the single link coupling to $J=1$ hereafter.

\begin{figure}
\centering
\includegraphics[width=0.9\columnwidth]{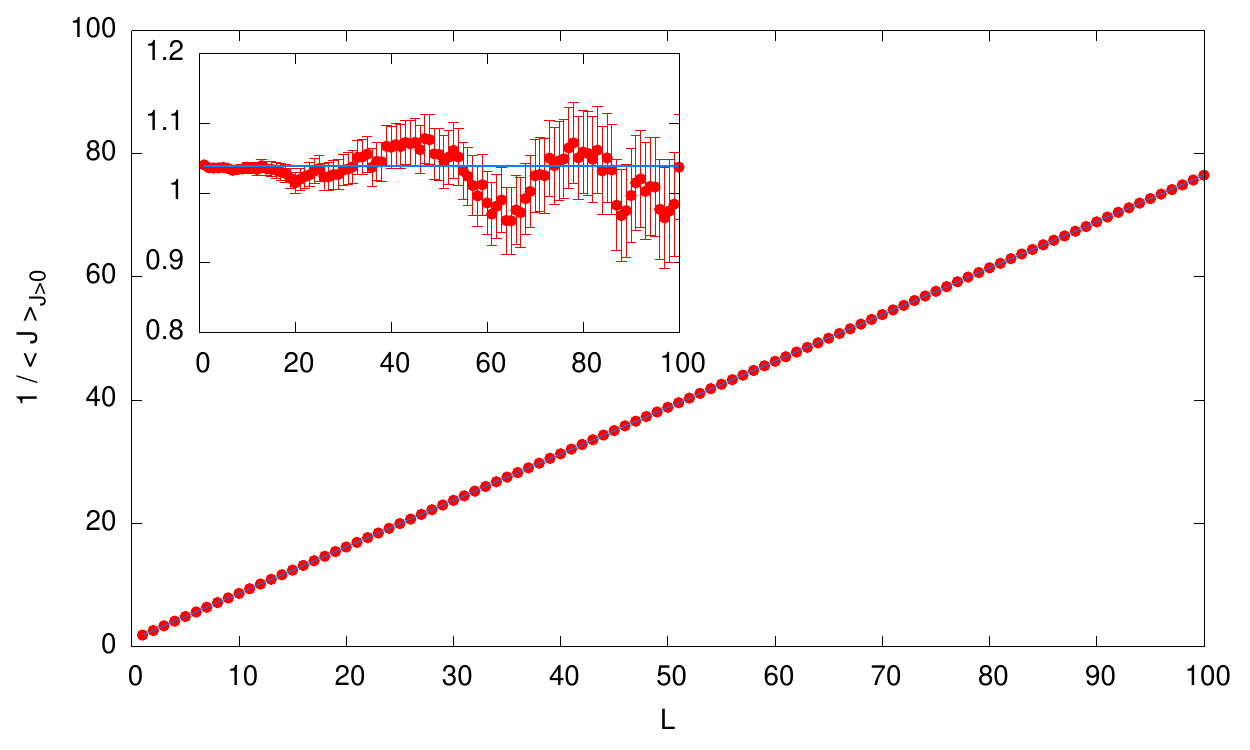}
\caption{\label{fig:rho}
The inverse of the mean effective coupling $\langle J_L \rangle_{J>0}$ scales linearly with the distance $L$ on the BL. The fit $\langle J \rangle_{J>0}^{-1} = 1.035(5) + 0.755(1) L$ interpolates well the data points, even at relatively small values of $L$, as can be appreciated in the inset, where we plot $\langle J \rangle_{J>0}^{-1} - 0.755 L$ versus $L$.}
\end{figure}

A possible numerical method has been already discussed in the previous sections and consists in evolving a population of triplets $(u_1,u_2,J)_L$ reweighted in a such a way that triplets with $J\neq 0$ do not decrease exponentially fast but remain constant in number: this trick allows to follow triplets with non-zero effective coupling for a long enough time to measure accurately the exponent $\rho$. As an example we show in Figure~\ref{fig:rho} the inverse of the mean effective coupling as a function of $L$, measured at the critical point $\sigma_{h,c}\simeq 1.037$ for $z=3$.

Although the fit shown in Figure \ref{fig:rho} is very good and provides an estimate to $\rho = 0.755(1)$ we have to remind that the reported uncertainty only represent the statistical error \emph{given the fitting function}. It is much more difficult to estimate the systematic error, that would depend --- among others --- on the corrections to the asymptotic scaling. For this reason we would be much more confident if we could derive an analytical expression for $\rho$.

Given that the effective couplings becomes very small even on the BL branches where they are non-zero, we would like to exploit this observation to better study the asymptotic distribution of cavity messages.
Let us consider the equations for updating the triplets reported in Table~\ref{table:BPrules} and let us rewrite it in a more explicit form, concentrating on the messages acting on the spin at distance $L$ (we ignore the messages arriving on the spin at the root). Schematically we have that, adding one new link, the messages change according to the following rules ($*$ messages are irrelevant in the present computation)
\begin{align}
&(*,u_L,J_L) + (0,0,1) \to (*,u_{L+1},J_{L+1})\\
&u_{L+1} = \hat{u}\big(h_{z-2}+u_L,J_L\big)\\
&J_{L+1} = \hat{J}\big(h_{z-2}+u_L,J_L\big)
\end{align}
where $h_{z-2}\sim Q_{z-2}$, i.e. $h_{z-2} =  h^R+ \sum_{i=1}^{z-2} u_i$ in distribution, and the functions are defined as follows
\begin{align}
\hat{u}(h,J) &= \sign(h) \left\{
\begin{array}{ll}
|h| & \text{if  } |h|<1-J\\
\frac{|h|+1-J}{2} & \text{if  } 1-J<|h|<1+J\\
1 & \text{if  } 1+J<|h|
\end{array}
\right.\label{eq:uhat}\\
\hat{J}(h,J) &= \left\{
\begin{array}{ll}
J & \text{if  } |h|<1-J\\
\frac{1+J-|h|}{2} & \text{if  } 1-J<|h|<1+J\\
0 & \text{if  } 1+J<|h|
\end{array}
\right.
\end{align}
From the above expressions we understand that during the evolution with probability $\mathbb{P}[1+J<|h|]$ the effective coupling becomes null, but we are interested in the complementary events, when the effective coupling remains non-zero.
With probability $\mathbb{P}[|h|<1-J]$ the coupling remains unaltered and with probability $\mathbb{P}[1-J<|h|<1+J]$ it decreases. We notice that the last event becomes very rare in the limit of small $J$, because the random variable $h$ has a continuous probability density function $P_h$ with no Dirac deltas in 1 or -1, so $\mathbb{P}[1-J<|h|<1+J]\simeq (P_h(1)+P_h(-1))2J$.

In practice in the large $L$ limit, when all effective couplings are very small, $J_L\ll 1$, the evolution proceeds essentially by keeping the $J_L$ constant until it jumps directly to $J_L=0$.

Let us move now to the analysis of the cavity messages $u_L$. Assuming we are in the large $L$ limit and all effective couplings are very small, we can work under the above hypothesis that the effective coupling stays constant in $L$ until it becomes null. So hereafter we fix $J_L=J$, where $J\ll 1$ is a small constant. We call $Q_J$ the probability distribution of the cavity messages on the branches where the effective coupling is fixed to $J$. From Eq.~[\ref{eq:uhat}] it is easy to derive that asymptotically $Q_J$ has support in $(-1+J,1-J)$ and satisfies the following equation
\begin{align}
&\lambda(J) Q_J(u') =\label{eq:QJ}\\
&=\mathbb{E} \int du Q_J(u) \delta\big(u'-(h_{z-2}+u)\big) \mathbb{I}(|u'|<1-J)\,,\nonumber
\end{align}
where $\mathbb{I}$ is the indicator function and the normalizing factor $\lambda(J)$ is given by
\begin{multline}
\lambda(J) = \int du Q_J(u) \mathbb{I}(|h_{z-2}+u|<1-J) =\\
= \mathbb{P}[|h_{z-2} + u^{(J)}|<1-J]\,,
\label{eq:lambdaJ}
\end{multline}
where $u^{(J)}\sim Q_J$.
In practice $\lambda(J)$ is the rate of survival of a non-zero effective coupling equal to $J$.
From this we can obtain the probability distribution of couplings in the $J\ll 1$ limit
\begin{multline}
\mathbb{P}[J_L=J] \propto \lambda(J)^L \simeq \big[\lambda(0)+\lambda'(0)J\big]^L \propto
\left(1+\frac{\lambda'(0)}{\lambda(0)}J\right)^L \propto\\
\propto \exp\left(\frac{\lambda'(0)}{\lambda(0)} J L\right) \propto \exp(-\rho J L) \implies
\rho = -\frac{\lambda'(0)}{\lambda(0)}\,.
\end{multline}
Given that we are mostly interested in studying the decay at the critical point it is worth reminding that at criticality $\lambda(0)=1/(z-1)$ holds and thus we have
\be
(z-1) \lambda(J) \simeq 1-\rho J \;\text{ for }\; J\ll 1\,.
\ee

\begin{figure}
\centering
\includegraphics[width=0.9\columnwidth]{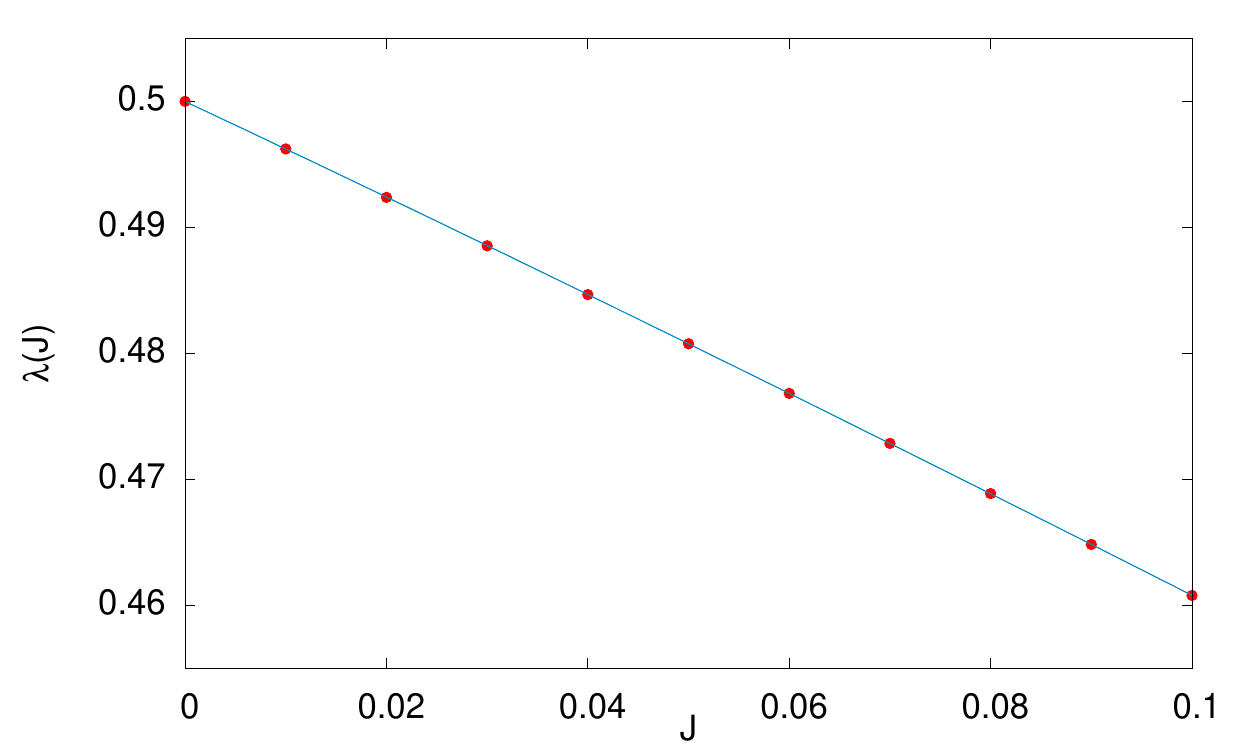}
\caption{\label{fig:lambdaJ}
$\lambda(J)$ computed at the critical field $\sigma_{h,c}=1.037$ for $z=3$. The curve is the best fit to the function $\lambda(J)=\lambda(0)+\lambda'(0)J+a J^2$. }
\end{figure}

In Figure \ref{fig:lambdaJ} we show data for $\lambda(J)$ computed at criticality for $z=3$ together with best interpolation via the following function
\begin{align}
    \lambda(J)=&\lambda(0)+\lambda'(0)J+a J^2 \label{eq:fit2}
\end{align}
The curve interpolates perfectly the data within the statistical uncertainties
and it returns an estimate $\rho=0.754(1)$, compatible with the numerical estimate coming 
from the triplets evolution described in previous sections.


\end{document}